\documentclass[iop]{emulateapj}
\tighten

\newcommand{\beq}    {\begin{equation}}
\newcommand{\eeq}    {\end{equation}}
\newcommand{\beqa}   {\begin{eqnarray}}
\newcommand{\eeqa}   {\end{eqnarray}}
\newcommand{\gsim}   {\mbox{$_>\atop^{\sim}$}}
\newcommand{\lsim}   {\mbox{$_<\atop^{\sim}$}}

\begin{document}

\title{ }

\title{Assessment of Models of Galactic Thermal Dust Emission Using {\it COBE}/FIRAS and {\it COBE}/DIRBE Observations} 

\author{
N. Odegard\altaffilmark{1},
A. Kogut\altaffilmark{2},
D. T. Chuss\altaffilmark{3},
N. J. Miller\altaffilmark{4,2}
}
\altaffiltext{1}{ADNET Systems, Inc., Code 665, NASA Goddard Space Flight Center, Greenbelt, MD 20771, USA; Nils.Odegard@nasa.gov.}
\altaffiltext{2}{Code 665, NASA Goddard Space Flight Center, Greenbelt, MD 20771, USA.}
\altaffiltext{3}{Department of Physics, Villanova University, 800 E. Lancaster Ave., Villanova, PA, 19085, USA.}
\altaffiltext{4}{Department of Physics and Astronomy, Johns Hopkins University, 3400 N. Charles St., Baltimore, MD, 21218, USA.}

\begin{abstract}
Accurate modeling of the spectrum of thermal dust emission at millimeter
wavelengths is important for improving the accuracy of foreground subtraction for CMB
measurements, for improving the accuracy with which the contributions of different
foreground emission components can be determined, and for improving our understanding
of dust composition and dust physics.  We fit four models of dust
emission to high Galactic latitude {\it COBE}/FIRAS and {\it COBE}/DIRBE observations 
from 3 millimeters to 100 $\mu$m and compare the quality of the fits.  We consider the 
two-level systems model because it provides a physically motivated explanation for
the observed long wavelength flattening of the dust spectrum and the anticorrelation
between emissivity index and dust temperature.  We consider the model of Finkbeiner, 
Davis, and Schlegel because it has been widely used for CMB studies, and the
generalized version of this model recently applied to {\it Planck} data by Meisner and
Finkbeiner. For comparison we 
have also fit a phenomenological model consisting of the sum of two graybody components.
We find that the two-graybody model gives the best fit and the FDS model gives a 
significantly poorer fit than the other models. The Meisner and Finkbeiner model
and the two-level systems model remain viable for use in Galactic foreground subtraction,
but the FIRAS data do not have sufficient signal-to-noise ratio to provide a strong test 
of the predicted spectrum at millimeter wavelengths.
\end{abstract}


\section{Introduction}

Diffuse thermal emission from interstellar dust is an important contributor to
Galactic foreground emission at millimeter to far-infrared wavelengths.
Accurate modeling of its spectrum at these wavelengths is important
for studies of the cosmic microwave background (CMB).  It is thought to be the
dominant contribution to the Galactic foreground at frequencies greater
than about 100 GHz (or wavelengths shorter than about 3 millimeters). Contributions
from free-free emission, synchrotron emission, and anomalous microwave emission
become increasingly important at lower frequencies, and the ratio
of CMB anisotropy to Galactic foreground anisotropy peaks at about 70 or 80 GHz
for high Galactic latitudes (Bennett et al. 2013, Planck Collaboration 2015 Results X 2015).
Improved accuracy in modeling the thermal dust spectrum at millimeter wavelengths
is important for improving the accuracy of foreground subtraction for CMB
measurements and would allow for more accurate separation of the
different foreground component spectra.  It would also help our understanding
of the composition of dust and the physics of dust emission at these wavelengths.

Observations of the spectrum of Galactic dust show excess emission at submillimeter and
millimeter wavelengths relative to a single temperature modified blackbody fit to the far infrared
spectrum (e.g. Wright et al. 1991, Reach et al. 1995, Finkbeiner, Davis, and Schlegel 1999,
Paradis et al. 2009, Planck Collaboration Int. XVII 2014, Planck Collaboration Int. XXII 2014, 
Meisner and Finkbeiner 2015).  This can be characterized
as a flattening of the dust emissivity spectral index $\beta$ determined from a modified blackbody fit
of the form
$I_{\nu} = \tau_{\nu_0} (\nu/\nu_0)^{\beta} B_{\nu}(T_d)$, where $\tau_{\nu_0}$ is the dust
optical depth at a reference frequency $\nu_0$, $B_{\nu}$ is the Planck function and $T_d$ is the dust
color temperature.  Planck Collaboration Int. XVII (2014) found a flattening $\beta_{FIR} - \beta_{mm}$
of 0.15 for H I correlated dust emission from the diffuse interstellar medium at high Galactic latitude
($b < -25\arcdeg$), with $\beta_{mm} =1.53 \pm 0.03$ over $100 \leq \nu \leq 353$ GHz. Planck Collaboration
Int. XXII (2014) found a similar result for emission correlated with a 353 GHz dust template at intermediate
Galactic latitude ($10\arcdeg < \vert b \vert < 60\arcdeg$).  Using a more recent {\it Planck} team internal
data release, Planck Collaboration Int. XXII (2015) found that band-to-band calibration uncertainties limit
the accuracy with which any flattening can be measured.

Some dust emission models account for the submillimeter/millimeter excess by incorporating a cold dust
component (e.g., Finkbeiner, Davis, and Schlegel 1999, Meisner and Finkbeiner 2015) or by using optical
properties of amorphous silicate
dust adjusted to better match the long wavelength observations (e.g., Draine and Li 2007, Compi{\`e}gne et al. 2011).
Other possible mechanisms include long wavelength emission due to low energy excitations in the
disordered internal structure of amorphous dust grains (Meny et al. 2007, Paradis et al. 2011),
emission from amorphous carbon dust with relatively flat emissivity at long wavelengths
(Compi{\`e}gne et al. 2011, Jones et al. 2013), and magnetic dipole emission from
magnetic grains or inclusions (Draine and Hensley 2013).

Many studies have found anticorrelations between dust emissivity spectral index and dust temperature
derived from modified blackbody fits at far infrared to millimeter wavelengths (e.g.,
Dupac et al. 2003, D{\'e}sert et al. 2008, Paradis et al. 2010, Veneziani et al. 2010,
Bracco et al. 2011,  Planck Collaboration Early XXIII 2011, Planck Collaboration Early XXV 2011,
Kelly et al. 2012, Liang et al. 2012, Juvela et al. 2013, Veneziani et al. 2013, Planck Collaboration Int. XVII 2014, 
Planck Collaboration 2013 Results XI 2014).  Results for the dispersions of $\beta$ and $T_d$ values
and the slope of their correlation vary among different studies; there are differences in sky
regions or source samples analysed, differences in datasets and wavelength coverages used, and differences
in analysis methods.
In many cases it appears that the anticorrelation cannot be explained purely as an effect of 
measurement uncertainties or fluctuations in the cosmic infrared background, suggesting that it may
be an intrinsic characteristic of dust emission.  Anticorrelation can also be caused by variations 
of dust temperature along the line of sight (e.g., Shetty et al. 1996).

Laboratory measurements of interstellar dust analogs at submillimeter to millimeter wavelengths have shown
the absorption coefficient spectral index to be anticorrelated with grain temperature for a number of
different amorphous grain materials
(Agladze et al. 1996, Mennella et al. 1998, Boudet et al. 2005, Coupeaud et al. 2011).
For crystalline materials that have been measured, temperature dependence is much weaker or not detected.
These results have been interpreted in terms of the two-level systems (TLS)
tunneling model for amorphous solids (Phillips 1972, Anderson et al. 1972, B{\"o}sch 1978).  The TLS model was
developed for interstellar dust by Meny et al. (2007) and Paradis et al. (2011).  It includes absorption
due to acoustic 
oscillations in a disordered charge distribution (DCD) and absorption due to resonant and relaxation
processes in a distribution of low energy two-level tunneling states.  The DCD absorption dominates
in the far infrared and is temperature independent.  Absorption due to hopping relaxation is 
the largest contributor at millimeter wavelengths and increases with increasing temperature.  The 
hopping relaxation produces long wavelength flattening of the emitted dust spectrum and anticorrelation
between emissivity spectral index and temperature.  Figure 1 shows the emissivity spectral index predicted
by the best-fit TLS model of Paradis et al. (2011) as a function of wavelength and dust temperature.

In this paper, we use {\it COBE}/FIRAS and {\it COBE}/DIRBE data to assess the quality of dust emission
spectrum predictions from different models in the 100 to 3000 GHz range (3 millimeters to 100 $\mu$m).
We consider two of the leading physically motivated models, the TLS model because it provides a natural 
explanation for both the long wavelength flattening of the spectrum and the $\beta - T_d$ anticorrelation, 
and the model of Finkbeiner, Davis, and Schlegel 1999 (hereafter FDS) because it has been widely used for CMB 
studies. This model incorporates two dust components with different power-law emissivities that are in
equilibrium with the interstellar radiation field.
Previous fits of these models have given a reduced $\chi^2$ of 1.85 for a fit of the FDS model
to FIRAS observations (Finkbeiner, Davis, and Schlegel 1999) and a reduced $\chi^2$ of 2.53 for a fit of 
the TLS model to FIRAS, {\it WMAP}, and Archeops observations (Paradis et al. 2011). Meisner and 
Finkbeiner (2015, hereafter MF) obtained a reduced $\chi^2$ of 1.33 from fitting a generalized version of 
the FDS model with more free parameters to {\it Planck} High Frequency Instrument (HFI) observations and 
DIRBE/{\it IRAS} 100 $\mu$m observations.

These results are not directly comparable because of differences in the analyses.  Here we make a direct
side-by-side comparison of the different models by applying the same analysis
method to a common dataset for a common sky region.  We compare the fit results for these physical models with
those for a purely phenomenological model consisting of the sum of two graybody components.  The paper is 
organized as follows.
In \S2 we describe the data sets used in the analysis.  \S3 describes our methods of calculating predictions
for the different dust models and fitting them to the data.  In \S4 we present the fit results and compare
them with results of previous studies.

\section{Data Sets}

We fit different models of interstellar dust emission to {\it COBE}/FIRAS
and {\it COBE}/DIRBE data from which estimated contributions of
the CMB, cosmic infrared background (CIB), zodiacal light, Galactic synchrotron
emission, and Galactic free-free emission have been subtracted.  These data sets 
provide the broad spectral coverage needed to determine the dust model parameters,
from the peak dust emission in the far infrared to the Rayleigh-Jeans tail at
millimeter wavelengths. Accurate calibration of the FIRAS data was established
by interspersing sky measurements with measurements of an external blackbody calibrator
that filled the FIRAS beam.

\subsection{{\it COBE}/FIRAS Data}

We use data from the final (pass 4) delivery of the FIRAS low spectral resolution
Destriped Sky Spectra, which cover 99\% of the sky over 213 frequency channels from 
68 GHz to 2911 GHz.  We subtract the CIB spectrum as determined by Fixsen et al. (1998), 
$I_{\nu}^{CIB} = 1.3 \times 10^{-5} (\nu/\nu_0)^{0.64} B_{\nu}(18.5$ K), where $\nu_0 = 3000$ GHz.
We subtract the publicly released FIRAS zodiacal light model (Brodd et al. 1997), 
which covers frequencies down to 600 GHz.  At lower frequencies where zodiacal
emission is not detectable in the FIRAS data, we subtract a power law extrapolation 
of the model.  We subtract Galactic free-free and synchrotron emission using results
from {\it WMAP} 9-year maximum entropy method foreground fitting (Bennett et al. 2013), 
extrapolated to FIRAS frequencies using a brightness temperature spectral index 
$\beta$ of -2.15 for free-free and -3.0 for synchrotron.  Free-free and synchrotron 
emission are negligible compared to the uncertainties in the FIRAS data for the sky 
regions and frequencies $\nu > 100$ GHz used in our analysis.  Anomalous 
microwave emission is also
thought to be negligible at these frequencies (Bennett et al. 2013, 
Planck Collaboration 2015 Results X 2015) and is not subtracted.  We subtract the
CMB dipole using the dipole amplitude and direction from {\it WMAP} (Hinshaw et al. 2009).
We used the data with these contributions subtracted to make an initial fit of each 
dust emission model as described below but with a CMB monopole component included.  
The CMB temperature obtained from the initial TLS model fit is 2.72765 K, and the
initial FDS and two-graybody fits gave results that agree within 0.00002 K.  
We subtract this 2.72765 K monopole to obtain the dataset used in our final fit for
each dust model.  This is consistent with the CMB temperature of $2.728 \pm 0.004$ K
reported by the FIRAS team from analysis of the pass 4 data (Fixsen et al. 1996).

Following FDS, we exclude from our analysis 
data below 100 GHz due to low signal-to-noise ratio for the dust emission, 
data above 2100 GHz due to calibration gain uncertainty $\gsim$ 3\%, and the
frequency channels listed in their appendix A due to spectral line emission
or problematic data.  Our analysis is more susceptible to spectral line contamination
than that of FDS, since we do not work with a dataset from which spectral 
line emission has been fit and subtracted.  To take care of line contamination 
over broader frequency ranges, we also exclude frequency channels 21, 32,
104, 106, 107, 135, 136, 141, and 150. (Following FDS, the numbering here starts at 
zero for the first channel, at 68 GHz.)  This leaves 111 channels in the range
100 GHz $< \nu < 2100$ GHz.

\subsection{{\it COBE}/DIRBE Data}

We use the final (pass 3B) delivery of the DIRBE Zodi-Subtracted Mission Average (ZSMA)
skymap at 240 $\mu$m. We subtract the 240 $\mu$m CIB as determined by Hauser et al. (1998).
Instead of using the DIRBE ZSMA map at 100 $\mu$m, we use the 100 $\mu$m dust emission map
of Schlegel, Finkbeiner, and Davis (1998, hereafter SFD) because this map is used in the
normalization of the FDS dust model. It was formed by combining DIRBE and {\it IRAS} 100 
$\mu$m maps, retaining the DIRBE gain and zero point calibration. The {\it IRAS} data
provide measurements of structure at angular scales smaller than the DIRBE beam ($\sim 0.7\arcdeg$).
A DIRBE 25 $\mu$m template fit was used to subtract zodiacal light and an H I template fit 
was used to set the zero level of Galactic dust emission. The 100 $\mu$m
and 240 $\mu$m maps are each smoothed to the FIRAS beam ($7\arcdeg$ FWHM) 
and then averaged over FIRAS pixels ($2.6\arcdeg$ on a side) in the {\it COBE}
quadrilateralized spherical cube projection.

\section{Analysis}

We fit each dust emission model to the Galactic dust spectra from the FIRAS and DIRBE data.  
To obtain results characteristic of the diffuse interstellar medium, we use the same mask as that
used by FDS to exclude the Galactic plane ($\vert b \vert < 7\arcdeg$), the Magellanic Clouds,
and H II regions in Orion and Ophiuchus.  The mask also excludes regions that may be affected
by FIRAS sidelobe contamination and excludes FIRAS pixels that were not observed or have low weight 
(weight less than 0.4).  The mask excludes 29\% of the sky and is shown superposed on a map
of FIRAS 1200 GHz (250 $\micron$) dust emission in Figure 2.
To reduce the time required for our fits to converge, we form mean dust spectra for six large regions
covering the sky, with
FIRAS pixels inside of the FDS mask excluded. Regions 1 through 4 are $90\arcdeg$ wide in Galactic longitude
and latitude, centered on the Galactic plane at $l = 0\arcdeg, 90\arcdeg, 180\arcdeg,$ and $270\arcdeg$. 
Regions 5 and 6 are the polar caps at $b < -45\arcdeg$ and $b > 45\arcdeg$.
For each region, a weighted mean spectrum is formed using pixel weighting based on the FIRAS 
detector noise and FIRAS destriper uncertainty $\beta$ (Brodd et al. 1997).  We fit each dust model
to the six region-averaged spectra. Table 1 lists the fit parameters that are global and those that
vary by region for each dust model.

\subsection{Evaluation of the TLS Dust Model}

The TLS model is a model of the far-infrared to millimeter wavelength emission from amorphous
dust grains, based on a solid-state physics description of the disordered internal structure
of the grains (Meny et al. 2007, Paradis et al 2011).  The model includes emission due to acoustic 
oscillations in the disordered charge distribution (DCD) in the grains and emission by a distribution
of localized two-level systems (TLS).  Three processes contribute
to the two-level systems emission: resonant tunneling, relaxation due to phonon-assisted tunneling,
and relaxation due to phonon-assisted hopping.  The model is parameterized by (1) the charge correlation
length $l_c$, which determines the wavelength where the spectral index of DCD absorption changes from
$\beta = 2$ to $\beta = 4$, (2) the amplitude $A$ of TLS absorption relative to DCD absorption, (3)
a parameter of tunneling states, $c_{\Delta}$, which determines the amplitude of hopping relaxation
relative to the other TLS processes, (4) the dust temperature $T_d$ for each sky region, and (5)
a scale factor $a$ that sets the overall amplitude of the spectrum for each region.  Assuming that 
$l_c$, $A$, and $c_{\Delta}$ are each constant over the sky, Paradis et al. (2011) determined best-fit
values for the model parameters using FIRAS and {\it WMAP} observations of the mean far infrared to millimeter
spectral energy distribution of the local interstellar medium, together with Archeops observations of 
far infrared to submillimeter spectral energy distributions for compact Galactic sources.  This showed
that the DCD process gives the largest contribution to the total emission at high frequencies and the
hopping relaxation process gives the largest contribution below about 170 GHz.  Paradis et al. (2012) fit the 
TLS model to 100-500 $\micron$ IRIS and {\it Herschel} Hi-GAL observations of the inner Galactic plane with 
$l_c$ and $A$ allowed to vary with position and $c_{\Delta}$ held fixed at the best-fit value from 
Paradis et al. (2011).  Values of $A$ were found to vary significantly in the plane, tracing variations in
excess 500 $\micron$ emissivity relative to a modified blackbody fit, and no significant variations
were found for $l_c$.  On average, $A$ was found to be significantly larger than the best-fit value
of Paradis et al. (2011).  This was interpreted as due to a higher degree of amorphization of the 
grains for the inner Galactic plane than for the solar neighborhood.  For our TLS model fitting
we assume that $l_c$, $A$, and $c_{\Delta}$ are each constant over the sky.  

We evaluate the predicted spectrum for a given set of parameters following the formulation of 
Paradis et al. (2011), except we calculate DCD absorption following B{\"o}sch (1978).  We found that 
this was necessary to reproduce the predicted DCD, resonant tunneling, and 
hopping relaxation spectra shown in Figure 2 of Paradis et al. (2011).  We have not been able to
reproduce the predicted spectra shown for tunneling relaxation, but the contribution of this process to the total
spectrum is negligible ($\lsim$ 1\% for the best-fit parameters of Paradis et al. and the frequency range
considered here) so this has no significant effect on our results. 

The predicted spectrum for a given region is calculated as 
\begin{equation}
I_\nu(\nu) = a \thinspace \alpha_{tot}(\nu,T_d,l_c,A,c_{\Delta}) \thinspace B_{\nu}(\nu,T_d),
\end{equation}
where a is the amplitude scale factor, $B_{\nu}$ is the Planck function, and
$\alpha_{tot}$ is the total absorption coefficient given by
\begin{equation}
\alpha_{tot} = \alpha_{DCD} + A \thinspace (\alpha_{res} + \alpha_{phon} + \alpha_{hop}).
\end{equation}
We calculate the DCD absorption coefficient following B{\"o}sch (1978) as
\begin{equation}
\alpha_{DCD} =\frac {2 (\epsilon + 2)^2} {3^3 \thinspace c \thinspace \sqrt{\epsilon} \thinspace v_t^3} \thinspace \langle \frac{q^2} {m} \rangle \thinspace \omega^2 \left[1 - \left(1 + \frac {\omega^2} {\omega_0^2}\right)^{-2}\right],
\end{equation}
where $\epsilon$ is the dielectric constant, $c$ is the speed of light, $v_t$ is the transverse sound speed 
in the material, $\langle \frac{q^2} {m} \rangle$ is the mean squared charge deviation per atomic mass,
and $\omega_0 = 2\pi v_t/l_c$ is the characteristic angular frequency where the spectral index of DCD absorption
changes from 2 to 4. We use $\epsilon$=2.6, $v_t=3 \times 10^5$ cm s$^{-1}$, and $\langle \frac{q^2} {m} \rangle=
6044$ erg cm g$^{-1}$ as used by B{\"o}sch (1978) for a soda-lime-silicate glass. 

We calculate the absorption coefficients for the TLS processes following Paradis et al. (2011).
For resonant tunneling,
\begin{equation}
\alpha_{res}=\frac {4 \pi^2} {3c \thinspace \sqrt{\epsilon}} \frac {(\epsilon + 2)^2} {9} \thinspace \omega \thinspace G(\omega) \thinspace \tanh(\hbar\omega/2kT_d),
\end{equation}
where $G(\omega)$ is the optical density of states, assumed to be constant at
$G(\omega) = G_0 = 1.4 \times 10^{-3}$ in cgs units. This is the value found by B{\"o}sch (1978) from fits to absorption measurements of a soda-silicate glass.
For phonon-assisted tunneling relaxation,
\begin{equation}
\alpha_{phon} = \frac {G_0} {3c \thinspace \sqrt{\epsilon}} \thinspace \frac {(\epsilon + 2)^2} {9} \thinspace \omega \thinspace F_2(\omega,T_d),
\end{equation}
where
\begin{equation}
F_2(\omega,T_d)= \frac {1} {2kT_d} \int_0^\infty \int^\infty_{\tau_1} \sqrt{1 - \frac {\tau_1} {\tau}} \enspace {\mathrm{sech}}^2 \left(\frac {E} {2kT_d} \right) \frac {\omega \thinspace d\tau dE} {1 + \omega^2\tau^2}.
\end{equation}
Here $\tau_1$ is the relaxation time defined by $\tau_1 = a E^{-3} \tanh (E/2kT_d)$, with $a$ set to $4.2 \times 10^{-56}$ erg$^3$ s as found by B{\"o}sch (1978) for soda-silica glass.
For phonon-assisted hopping relaxation,
\begin{equation}
\alpha_{hop} = \frac{8\pi} {3c \thinspace \sqrt{\epsilon}} \frac {(\epsilon + 2)^2} {9} \thinspace G_0 (c_{\Delta} + \ln T_d) \int_0^\infty dV P(V) \thinspace \frac {\omega^2 \tau} {1 +  \omega^2 \tau^2},
\end{equation}
where $V$ is the TLS potential energy barrier height and $\tau$ is the relaxation time defined by $\tau = \tau_0$ exp$(V/T_d)$ with
$\tau_0 \simeq 1 \times 10^{-13}$ s from B{\"o}sch (1978).
The tunneling states parameter $c_\Delta$ is given by
\begin{equation}
c_{\Delta} = \ln \frac {k} {\Delta_0^{min}} - 0.441
\end{equation}
(Meny et al. 2007) where $\Delta_0^{min}$ is the tunnel splitting. The experimental upper limit for $\Delta_0^{min}$
corresponds to $c_{\Delta} \gsim 5.8$.   The distribution of
TLS barrier height $P(V)$ is approximated as
\begin{equation}
P(V) = \frac {\mathrm{exp}(-(V-V_m)^2/V_0^2)} {V_0 \sqrt{\pi}}
\end{equation}
for $V > V_{min}$ and $P(V) = 0$ for $V < V_{min}$, with $V_{min} = 50$ K, $V_m = 550$ K, and $V_0 = 410$ K from ultrasonic measurements (B{\"o}sch 1978).

Many of the physical parameters adopted for the absorption coefficient calculations are dependent on material characteristics and
are not well known for interstellar dust, so the model includes the fit parameter $A$ and also treats $c_{\Delta}$ as a fit parameter.

To obtain model predictions that can be compared with the quoted observations in the DIRBE bands, 
we apply DIRBE color correction factors based on the predicted spectrum to the model predictions at 
100 $\mu$m and 240 $\mu$m.\footnote{DIRBE provided maps of monochromatic brightness at the nominal
wavelength for each band assuming that the spectrum $\nu I_\nu$ is constant across the band.  For a model
spectrum that differs from this, a color correction calculated from the model spectrum and the DIRBE bandpass 
response function is needed.}

\subsection{Evaluation of the FDS Dust Model}

The best fit model of FDS (their model 8) consists of two dust components, each of which has a power-law 
emissivity and is in equilibrium with the interstellar radiation field.  FDS tentatively identified 
the colder dust component with amorphous silicate grains and the warmer component
with carbonaceous grains.  The model is parameterized by the
emissivity indices of the cold and warm components, $\alpha_1$ and $\alpha_2$, the fraction of the total
emitted power that comes from the cold component, $f_1$, and the ratio $q_1/q_2$, where $q_i$ is
the ratio of the far infrared emission cross section at a reference frequency $\nu_0 = 3000$ GHz
to the effective UV-optical absorption cross section for dust component $i$. Each of these parameters
is constant over the sky.  For a given set of model parameters, a DIRBE 100 $\mu$m/240 $\mu$m band ratio
map is used to determine the temperatures of the two components and the shape of the model spectrum for
each map pixel.  The SFD 100 $\mu$m dust emission map 
determines the amplitude of the model spectrum.  FDS determined values for the model parameters by fitting
the model to correlation slopes between FIRAS channel maps and the SFD 100 $\mu$m dust map for the
diffuse emission region outside of the mask shown in Figure 2.

We evaluate the FDS model as follows for a given set of model parameters.  FDS used the requirement that
each dust component is in equilibrium with the interstellar radiation field to obtain a relation 
between the dust temperatures of the components, $T_1$ and $T_2$,
\begin{equation}
T_1^{4+\alpha_1} = \frac {q_2 Z(\alpha_2)} {q_1 Z(\alpha_1)} \left(\frac {h \nu_0} {k_B} \right)^{\alpha_1 - \alpha_2} T_2^{4+\alpha_2} ,
\end{equation}
where 
\begin{equation}
Z(\alpha) \equiv \int_0^\infty \frac {x^{3+\alpha}} {e^x - 1} dx = \zeta(4+\alpha) \Gamma(4+\alpha) .
\end{equation}
We set up a grid of $T_2$ values varying from 10 to 31 K in steps of 0.1 K and calculate the
corresponding values of $T_1$. 
\pagebreak
We then calculate values of the DIRBE 100 $\mu$m/240 $\mu$m band ratio
$R$ at the grid points using
\begin{widetext}
\begin{equation}
R = \frac {K_{100}(\alpha_1,T_1) f_1 \left(\frac {q_1} {q_2} \right) B_{\nu}(\nu_{100},T_1) + 
           K_{100}(\alpha_2,T_2) (1-f_1) B_{\nu}(\nu_{100},T_2)}
 {K_{240}(\alpha_1,T_1) f_1 \left(\frac {q_1} {q_2} \right) \left(\frac {\nu_{240}} {\nu_{100}} \right)^{\alpha_1} B_{\nu}(\nu_{240},T_1) +
  K_{240}(\alpha_2,T_2) (1-f_1) \left(\frac {\nu_{240}} {\nu_{100}} \right)^{\alpha_2} B_{\nu}(\nu_{240},T_2)} ,
\end{equation}
\end{widetext}
where $K_{100}$ and $K_{240}$ are color correction factors for the DIRBE 100 $\mu$m and 240 $\mu$m bands, and
$\nu_{100}$ and $\nu_{240}$ are the nominal frequencies for the DIRBE 100 $\mu$m and 240 $\mu$m bands.

We use these results with the observed DIRBE 100 $\mu$m/240 $\mu$m ratio map from FDS to determine
$T_1$ and $T_2$ for each HEALPix pixel at $N_\mathrm{side} = 512$ (pixel size $0.11 \arcdeg$).
The shape of the predicted spectrum $Y(\nu)$, normalized to the brightness in the DIRBE 100 $\mu$m band, 
is then calculated for each pixel using
\begin{widetext}
\begin{equation}
Y(\nu) = \frac {f_1 \left(\frac {q_1} {q_2} \right) \left(\frac {\nu} {\nu_{100}} \right)^{\alpha_1} B_{\nu}(\nu,T_1) +
            (1-f_1) \left(\frac {\nu} {\nu_{100}} \right)^{\alpha_2} B_{\nu}(\nu,T_2)}
               {K_{100}(\alpha_1,T_1) f_1 \left(\frac {q_1} {q_2} \right) B_{\nu}(\nu_{100},T_1) +
                K_{100}(\alpha_2,T_2) (1-f_1) B_{\nu}(\nu_{100},T_2)},
\end{equation}
\end{widetext}
and the predicted spectrum is obtained by multiplying the spectral shape by the brightness from
the 100 $\mu$m dust emission map made from DIRBE and {\it IRAS} data by SFD.

We calculate DIRBE 100 $\mu$m and 240 $\mu$m color correction factors from the model parameters
and dust component temperatures, and apply them to the predicted brightness values at $\nu_{100}$ and 
$\nu_{240}$ to obtain values that can be compared with the quoted DIRBE observations.  The model predictions
at each FIRAS frequency and the color-corrected model predictions for each DIRBE band are then smoothed to the 
FIRAS beam, averaged over FIRAS pixels, and averaged over sky regions using the same pixel weighting
as was used for the observed data.

We note some small differences between our analysis and that of FDS.
FDS applied a 1\% gain recalibration to the FIRAS data to correct for an inconsistency they found 
between the FIRAS data and the DIRBE 240 $\mu$m data.  We do not apply any recalibration, but discuss the
effect of a possible calibration difference on our results in \S4. The SFD 100 $\mu$m dust emission map used
by FDS is at $0.7 \arcdeg$ resolution and includes point sources. We use the version at $6 \arcmin$ resolution
with point sources subtracted.  The contribution of stars and galaxies is $\lsim $2\% of the Galactic dust
emission for the region used in our analysis.
There are also small differences in the DIRBE color correction used in the two analyses. The DIRBE Explanatory Supplement
tabulates color correction values for modified blackbody spectra over the parameter ranges $0 \leq \beta \leq 2.0$
and $10 \leq T \leq  20000$ K.  FDS fit a function to the tabulated values to interpolate and extrapolate in $T$.
We use DIRBE bandpass response functions to calculate the color correction for a given ($\beta,T$).  For the
100 $\mu$m color correction factor, the method used by FDS generally agrees with the direct calculation within 1\% 
but it underestimates the color correction below 10 K, by as much as 4\% for $\beta$ = 1.67, $T$=9.0 K.
FDS did not publish their color correction fit coefficients for the 240 $\mu$m band.

\subsection{Evaluation of the MF Dust Model}

We fit the generalized version of the FDS model as formulated by Meisner and Finkbeiner (2015). This version 
treats one of the dust temperatures, $T_2$, and the spectrum amplitude as fit parameters for each region,
instead of setting them using the DIRBE 100 $\mu$m/240 $\mu$m ratio map and the SFD 100 $\mu$m
dust map.
It retains the four global fit parameters of the FDS model ($\alpha_1, \alpha_2, f_1, q_1/q_2$) and the
relation between the dust temperatures
$T_1$ and $T_2$ for each region (equation 10). 
Meisner and Finkbeiner (2015) fit the model to correlation slopes of {\it Planck} HFI and DIRBE data relative to
{\it Planck} 857 GHz data for a high Galactic latitude region to determine values for the global parameters, and then 
made a fit to {\it Planck} 217, 353, 545, and 857 GHz maps and the SFD 100 $\mu$m map at 6.1 arcminute resolution
over the entire sky with the global parameter values held fixed. For diffuse sky regions, they found that the model
predictions agree with {\it Planck} 100, 143, and 217 GHz observations within 2.2\% on average, and there is similar
quality agreement with predictions of the {\it Planck} team all-sky thermal dust 
emission model (Planck Collaboration 2013 Results XI 2014) from 353 to 3000 GHz.

The predicted spectrum for this model is given by

\begin{equation}
I_{\nu}(\nu) = \tau \thinspace \frac{\left[f_1 \frac{q_1}{q_2} \left(\frac{\nu}{\nu_0}\right)^{\alpha_1} B_{\nu}(\nu,T_1) + f_2 \left(\frac{\nu}{\nu_0}\right)^{\alpha_2} B_{\nu}(\nu,T_2)\right]}{\left[f_1 \frac{q_1}{q_2} + f_2\right]}
\end{equation}
where $\tau$ is the total dust optical depth at a reference frequency $\nu_0$ and $f_2 = 1-f_1$.
We apply DIRBE color correction factors
to the 100 $\mu$m and 240 $\mu$m model predictions to obtain values that can be compared with the quoted
DIRBE observations.

\subsection{Evaluation of the Two-Graybody Model}

We also fit a model consisting of a sum of two modified blackbody components,
\begin{equation}
I_{\nu}(\nu) = \tau_1 \left(\frac{\nu}{\nu_0}\right)^{\alpha_1} B_{\nu}(\nu,T_1) + \tau_2 \left(\frac{\nu}{\nu_0}\right)^{\alpha_2} B_{\nu}(\nu,T_2).
\end{equation}
Here the parameters $\tau_1$ and $\tau_2$ are optical depths of the two components at a reference frequency $\nu_0$,
which we take to be 1000 GHz, $\alpha_1$ and $\alpha_2$ are emissivity indices, and $T_1$ and $T_2$ are
color temperatures.  This is a purely phenomenological model and parameter values from fitting it do not 
necessarily have any direct physical interpretation.  The emissivity indices are taken to be constant over the sky.  The 
temperatures and optical depths may vary from region to region.  We have found large degeneracies between parameters
if all of them are allowed to be free,
so we fix $\alpha_1$ at 1.7 (close to the best-fit value for component 1 in the FDS model) and solve for the 
other parameters.  We apply DIRBE color correction factors
to the 100 $\mu$m and 240 $\mu$m model predictions to obtain values that can be compared with the quoted
DIRBE observations.

\subsection{Fit Method}
For each model, initial parameter values and parameter covariance matrix are obtained using the Levenberg-Marquardt
method to minimize $\chi^2$ calculated using the diagonal of the noise covariance matrix. Final parameter
values are obtained using the downhill simplex method of Nelder and Mead (1965) to minimize $\chi^2$ 
calculated using the full noise covariance matrix,
\begin{equation}
\chi^2 = (\mathbf{I - M})^T \mathbf{C^{-1}} (\mathbf{I - M}),
\end{equation}
where \textbf{I} is a vector of the observations for all 6 regions and 113 frequencies, \textbf{M} is a 
correponding vector of model predictions, and $\mathbf{C^{-1}}$ is the inverse of the noise covariance matrix.

We include the following FIRAS uncertainties in the noise covariance matrix: detector noise, destriper
uncertainty ($\beta$), bolometer model gain uncertainty (JCJ gain), calibration model emissivity gain 
uncertainty (PEP gain), and internal calibrator temperature uncertainty (PUP). These are calculated as 
described in section 7 of the FIRAS Explanatory Supplement (Brodd et al. 1997).  Uncertainty in the
absolute temperature scale of the FIRAS external calibrator (PTP) is not included, since it is only 
important in determination of the absolute temperature of the CMB (Brodd et al. 1997).  The DIRBE 
uncertainties included are absolute calibration gain uncertainty (13.5\% at 100 $\mu$m and 11.6\% at 
240 $\mu$m) and absolute calibration offset uncertainty. The DIRBE 240 $\mu$m uncertainty is much
larger than the FIRAS uncertainties at nearby wavelengths, so the DIRBE 240 $\mu$m data has little effect
on our fit results.

\section{Results}

Results of our fits of the TLS, FDS, MF, and two-graybody models to the Galactic dust spectra are shown
in Figure 3 for a selected region.  For the TLS model fit, the DCD process contributes most of the emission 
above 250 to 300 GHz and hopping relaxation gives the largest contribution at lower frequencies.  For 
the other model fits, the cold dust component gives the largest contribution below about 500 GHz.

Global fit parameters from the model fits are listed in Table 2 and regional fit parameters or regional 
derived parameters are listed in Table 3.  For the FDS model, our fit parameter values are 
in good agreement (within 2\%) with those obtained by FDS for the same sky region from fitting the model
to correlation slopes between FIRAS data and 100 $\mu$m dust emission. For the MF model, our parameter
values are mostly consistent with those obtained by MF for a different, higher Galactic latitude region 
from fitting the model
to correlation slopes of {\it Planck} HFI and DIRBE data relative to HFI 857 GHz data ($\alpha_1 = 1.63$,
$\alpha_2 = 2.82$, $f_1 = 0.0485$, $q_1/q_2 = 8.22$, $T_2=15.70$). However, our $\alpha_1$ value is 
significantly flatter than that obtained by MF.  For the TLS model, our global parameter values
are consistent within the uncertainties with the 'standard' values of Paradis et al. 2011 ($l_c = 
13.40 \pm 1.49$ nm, $A = 5.81 \pm 0.09$, $c_{\Delta} = 475 \pm 20$) from 
fitting FIRAS and {\it WMAP} observations of the diffuse interstellar medium and Archeops observations
of compact Galactic sources.  Our region-averaged $T_{dust}$ is $18.9 \pm 0.3$ K, compared
with the Paradis et al. standard value $17.26 \pm 0.02$ K, or $17.53 \pm 0.02$ K from their fit
with the compact source observations excluded. It is not clear if the discrepancy is significant
since there are differences in sky coverage.
Paradis et al. used FIRAS and {\it WMAP} data for 14\% of the sky where $\vert b \vert > 6\arcdeg$ and
I(240 $\mu$m) $> 18$ MJy sr$^{-1}$, and we used data for 71\% of the sky where $\vert b \vert > 7\arcdeg$
(Figure 2).

We note that the amplitude of the hopping relaxation contribution from our fitting is more accurately determined
than would be expected from the large uncertainty in $c_{\Delta}$ given in Table 2.  There is a strong degeneracy
between $c_{\Delta}$ and $A$, and the hopping relaxation amplitude is proportional to their product.
We estimate the uncertainty in this amplitude to be about 17\% from fitting the model with a 
parameterization that uses a hopping relaxation amplitude parameter instead of $c_{\Delta}$.

Figure 4 shows a comparison of dust spectra and model fits averaged over the six regions, together with region-averaged
fit residuals (dust spectrum minus model fit) and fractional fit residuals.  The MF and two-graybody models 
give a reasonably good fit over the full frequency range, but the TLS and FDS model fits each overpredict the average
dust spectrum above about 700 GHz. Figure 5 shows the fractional residual of the different model fits for each region.
We note that the difference between the 240 $\mu$m DIRBE data points and neighboring FIRAS data points in Figures 4 and 5
is mostly due to calibration differences (Fixsen et al. 1997). The DIRBE 240 $\mu$m uncertainty is much
larger than the FIRAS uncertainties at nearby wavelengths, so the DIRBE 240 $\mu$m data points have little effect
on our fit results. The FDS model fit passes straight through the 100 $\mu$m data points because the same 
SFD 100 $\mu$m map is used for the model normalization and the data.  The FDS model exceeds the 240 $\mu$m DIRBE data
points because of differences between the FDS 100 $\mu$m/240 $\mu$m ratio map and the 100 $\mu$m/240 $\mu$m ratio
in the data we are using. This must be due to differences in the methods used in forming the zodiacal light and CIB
subtracted map at 240 $\mu$m.

Measures of the quality of the model fits are listed in Table 2.  
Column 3 lists the $\chi^2$ of the fits from equation 15 and column 5 lists the probability of
a larger $\chi^2$ value for the number of degrees of freedom $\nu$ given in column 4. The FDS model is strongly
disfavored compared to the TLS model, and both are worse than the MF and two-graybody models.
The two-graybody model gives the best fit because it has the most free parameters for each region.
The probability of a larger $\chi^2$ for the two-graybody model is close to unity, indicating that our
noise covariance matrix probably overestimates the actual uncertainties somewhat. To get a reduced $\chi^2$ of
unity for the two-graybody fit, the uncertainties would need to be decreased by 11\%. This would give 
$\chi^{2}_{\nu}$ values of 1.29, 1.12, and 3.84 and probability of a larger $\chi^2$ of $7 \times 10^{-7}$,
0.017, and $2 \times 10^{-266}$ for the TLS, MF, and FDS fits, respectively.

The TLS, MF, and two-graybody models give acceptable quality fits but the TLS model has very different behavior
below 300 GHz where the signal-to-noise ratio of the FIRAS data worsens (Figure 3). The emissivity index
of the TLS spectrum continues to flatten with decreasing frequency below 300 GHz. {\it Planck} data
may help to better constrain the different models and differentiate between them at low frequency, or data
from future experiments may be needed.  From an analysis of a 2014 data release internal to the {\it Planck}
team (release DX11d) together with DIRBE 100 $\mu$m
data, Planck Collaboration Int. XXII (2015) found that the band-to-band calibration accuracy is not good 
enough to either confirm or rule out the flattening predicted by the TLS model.  The calibration accuracy 
of the 2015 data release is improved (Planck Collaboration 2015 Results VIII 2015), but the improvement for
the 545 GHz and 857 GHz bands is small (5\% uncertainty for the 2015 release compared to 7\% for DX11d).

In principle, the FDS model might be expected to be better than the others for fitting dust spectra from data 
averaged over large sky regions since its use of the 100 $\mu$m/240 $\mu$m ratio map and the 100 $\mu$m map allows
it to account for variations of dust temperatures and spectrum amplitude on angular scales as small as 1 degree
and 6 arcminutes, respectively. However, we find that it gives a poor quality fit because it does not reproduce
the large-scale spatial distribution of dust emission as well as the other models do.  All of its fit parameters 
are global, while the other models are free to adjust the amplitude and dust temperature(s) for each region. For
the high Galactic latitude regions, Figure 5 shows that the FDS model overpredicts the observations by 10 to 20\%
over much of the frequency range. Figure 6 shows the residual of the FDS fit calculated at FIRAS resolution 
($\sim7\arcdeg$ FWHM) and averaged over seven channels centered at 1200 GHz (250 $\mu$m).  Trends with both
Galactic latitude and ecliptic latitude are present, and are also seen in the latitude profiles from the 
residual map shown in Figure 7. 

The ecliptic latitude
dependence is probably due to differences between the FIRAS team zodiacal light subtraction, which is based on
interpolation and extrapolation of the Kelsall et al. (1998) zodiacal light model to FIRAS frequencies, and the 
zodiacal light subtraction used in forming the SFD 100 $\mu$m map and the FDS 100 $\mu$m/240 $\mu$m ratio map,
which is based on fits to DIRBE 100 $\mu$m and 240 $\mu$m maps using a DIRBE 25 $\mu$m map as a zodiacal light
template and an H I map as a Galactic dust emission template.

The Galactic latitude residual profile is consistent with the corresponding Galactic latitude dust emission
profile scaled by 2.8\%. This suggests that the Galactic latitude dependence may be
a result of inconsistency between the SFD 100 $\mu$m map and the FIRAS data due to differences between the DIRBE
and FIRAS calibrations. The factor of 2.8\% is within the DIRBE 100 $\mu$m absolute calibration gain uncertainty of 13.5\%
and is within the range of DIRBE-FIRAS 100 $\mu$m cross-calibration results obtained by Fixsen et al.
(1997). Other possible causes for the Galactic latitude dependence include variations in one or more of the model
parameters that are taken to be global, or possible systematic errors in the FDS 100 $\mu$m/240 $\mu$m
ratio map at high Galactic latitudes. Such an error in the ratio map could result from errors in the zero levels
of the maps it was formed from, due to error in calibration, CIB subtraction, or zodiacal light subtraction. 
For example, the 1$\sigma$ uncertainty of the
240 $\mu$m CIB from Hauser et al. (1998), which includes calibration and zodiacal light subtraction uncertainties,
is about 25\% of the mean brightness of the CIB subtracted, zodiacal light subtracted 240 $\mu$m data at the Galactic
poles.

To see if the FDS model can be modified to reduce the fit residuals, we have tried fitting a modified version of
the model that uses the SFD 100 $\mu$m map rescaled by 1.028 and a 100 $\mu$m/240 $\mu$m ratio map 
calculated from the {\it Planck} team all-sky thermal dust emission model (Planck Collaboration 2013 Results XI 2014)
in place of the FDS ratio map.  The {\it Planck} model was determined
from fits of a modified blackbody to {\it Planck} 353 GHz, 545 GHz, and 857 GHz data with the {\it Planck} 
zodiacal emission model subtracted and 100 $\mu$m data from a combination of the SFD map at angular scales 
larger than $30\arcmin$
and the IRIS map (Miville-Deschenes and Lagache 2005) at smaller scales.  The zero level of each input map was
set by correlation with Galactic H I 21-cm line data at low column densities.  We used the {\it Planck} model
fit parameters to calculate maps of quoted brightness for the DIRBE 100 $\mu$m and 
240 $\mu$m bands, formed the ratio map, and smoothed it to 1 degree FWHM for use in fitting the modified
FDS model.  This ratio map has better signal-to-noise ratio than the FDS ratio map, and the maps 
also differ due to differences in zero levels of the input maps and differences in the zodiacal light subtraction
for the 240 $\mu$m band.  The ratio map from the {\it Planck} model is systematically larger than the FDS ratio map
at high Galactic latitudes, reflecting the increase of the {\it Planck} model dust temperature with latitude.
Figure 7 shows that the modified FDS model fit leads to improvement in both the Galactic latitude profile
and the ecliptic latitude profile of the 1200 GHz residual map. The reduced $\chi^2$ of the fit is 1.11,
much improved relative to the FDS model fit. This supports the proposition that the FDS model shortcomings in
reproducing the spatial distribution of dust emission are mostly due to inconsistency between the DIRBE and FIRAS
calibrations and inconsistency between the zodiacal light subtraction, CIB subtraction, and zero-level calibration
used for the 100 $\mu$m and 240 $\mu$m maps and that used for the FIRAS data.

Our fitting does not take into account systematic uncertainties associated with the subtraction of the 
contributions other than Galactic dust emission from the data.  The largest of these uncertainties is in 
subtraction of the CIB. We have investigated its effect by making fits where we have increased or 
decreased the amount of CIB removed by one sigma at all frequencies.
We used the one sigma CIB amplitude uncertainties in the FIRAS channels from Fixsen et al. (1998) and the uncertainties
in the DIRBE bands from Hauser et al. (1998).  The uncertainties are as large as 0.27 MJy sr$^{-1}$ at 1400 GHz.  
Fitting for the case where the adopted CIB was increased gave significantly poorer fits for the TLS
 and FDS models and only slightly poorer fits for the MF and two graybody models, with $\chi^{2}_{\nu}$ of
1.71, 7.82, 0.95, and 0.83 for the TLS, FDS, MF, and two graybody fits, respectively. Fitting for the case
where the CIB was decreased gave an improved fit for the FDS model and slightly poorer fits for the other models,
with $\chi^{2}_{\nu}$ of 1.06, 1.60, 0.93, and 0.85 for the TLS, FDS, MF, and two graybody fits, respectively.
In both cases the FDS fit is still significantly worse than the TLS fit, and the dependence of the FDS fit 
residuals on Galactic latitude and ecliptic latitude is similar to that shown in Figures 6 and 7 for our
original FDS fit.
 
\section{Summary and Conclusions}

We have assessed the quality of fits of the two-level systems, Finkbeiner, Davis, and Schlegel (1999), 
and Meisner and Finkbeiner (2015) models of
thermal dust emission to {\it COBE}/FIRAS and {\it COBE}/DIRBE observations of high latitude Galactic dust emission 
from 100 to 3000 GHz. This is the first side-by-side comparison of these models using a common 
dataset and a common sky region. For comparison we have also fit a phenomenological model consisting of the
sum of two graybody components.

The $\chi^2$ of the fits show that the FDS model is strongly disfavored compared to the other models. 
It does not reproduce the spatial distribution of dust emission at submillimeter wavelengths 
as well as the other models do. This appears to be mostly due to inconsistency between the DIRBE and FIRAS
calibrations and inconsistency between the zodiacal light subtraction, CIB subtraction, and zero-level calibration
used for the 100 $\mu$m and 240 $\mu$m maps in the FDS model and that used for the FIRAS data.  We found
big improvement in $\chi^2$ for a modified version of the 
FDS model that accounts for the apparent inconsistency between the DIRBE and FIRAS calibrations and
uses an input 100 $\mu$m/240 $\mu$m ratio map calculated from the {\it Planck} team all-sky
thermal dust emission model.  Variations of the fit residuals with ecliptic latitude and Galactic latitude
are significantly reduced.  However, the fit is still not quite as good as the fits of the other models,
which have freedom to adjust the spectrum amplitude and dust temperature(s) for each sky region.

The Meisner and Finkbeiner model and the TLS model give acceptable quality fits and remain viable
physical models for Galactic foreground
subtraction in CMB studies.  However, the FIRAS data do not provide a strong test of the predicted
spectrum below about 300 GHz due to decreasing signal-to-noise ratio. At these frequencies, the TLS model 
predicts progressive flattening of the spectrum with decreasing frequency that is not predicted by the
two-graybody, FDS, or MF models. {\it Planck} data may help to better 
constrain different dust models and differentiate between them at low frequency, as will
future experiments with large numbers of channels between about 60 and 3000 GHz such as the proposed 
Primordial Inflation Explorer (PIXIE) mission.

\bigskip

We thank D. Finkbeiner for providing the mask used by FDS, D. Fixsen for helpful discussions about FIRAS
data, and D. Paradis for help with questions about the TLS model. We thank the referee for helpful comments.
We acknowledge the use of the Legacy Archive for Microwave Background Data Analysis (LAMBDA), part of the
High Energy Astrophysics Science Archive Center (HEASARC). HEASARC/LAMBDA is a service of the Astrophysics
Science Division at the NASA Goddard Space Flight Center.

\newpage

\begin{figure}
\figurenum{1}
\epsscale{1.00}
\plotone{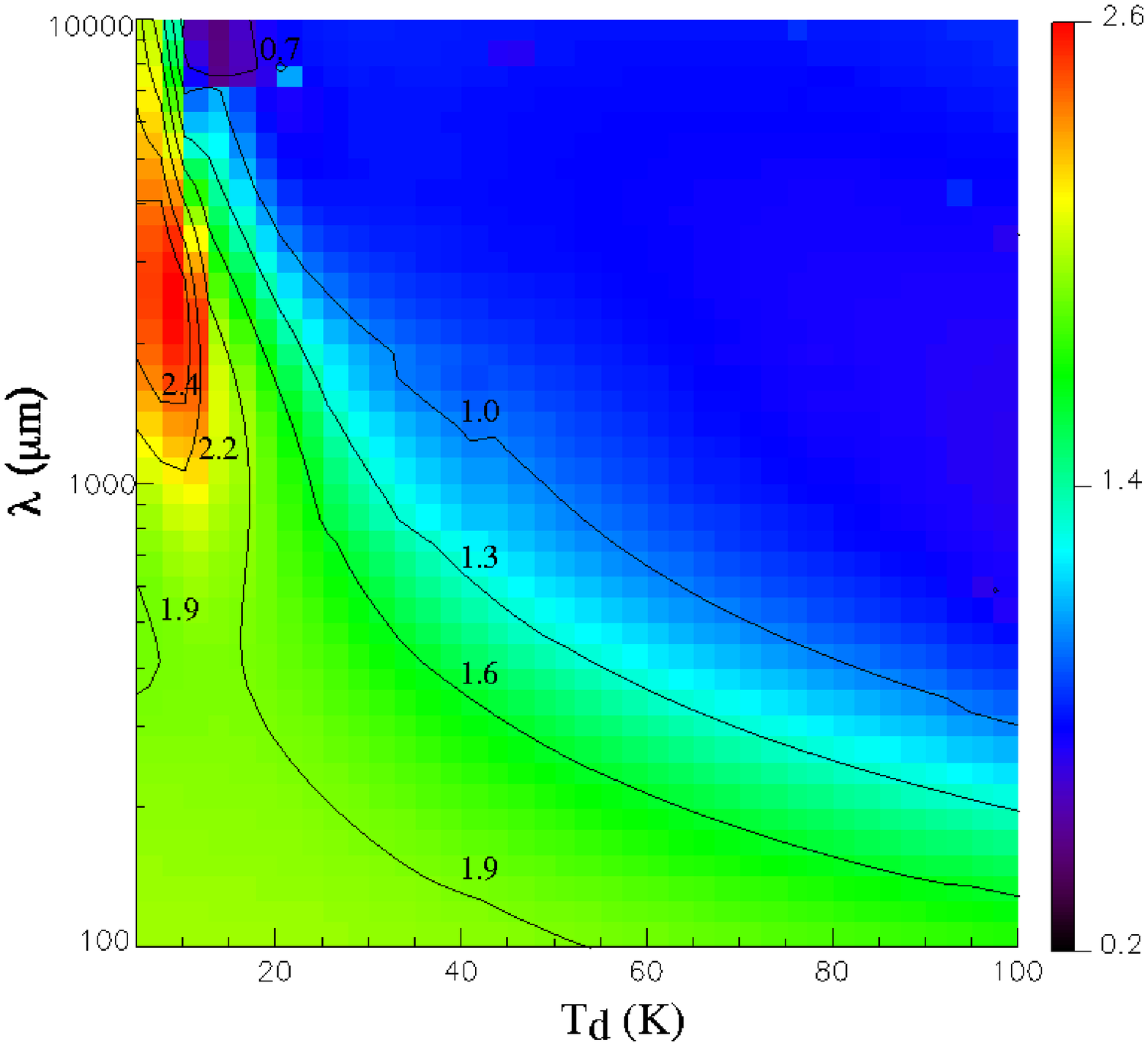}
\caption{The dust emissivity spectral index predicted by the two-level systems model as a function of
wavelength and dust temperature.  This figure is from Paradis et al. (2011) and is based on their best-fit
model parameters.  The model provides a natural explanation for both the observed long wavelength flattening of
the spectrum and the anticorrelation between emissivity spectral index and dust temperature.
Credit: Paradis et al., A\&A, 534, 118, 2011, reproduced with permission \copyright ESO. 
}
\end{figure}

\newpage

\begin{figure}
\figurenum{2}
\epsscale{1.00}
\includegraphics[scale=0.75,angle=90]{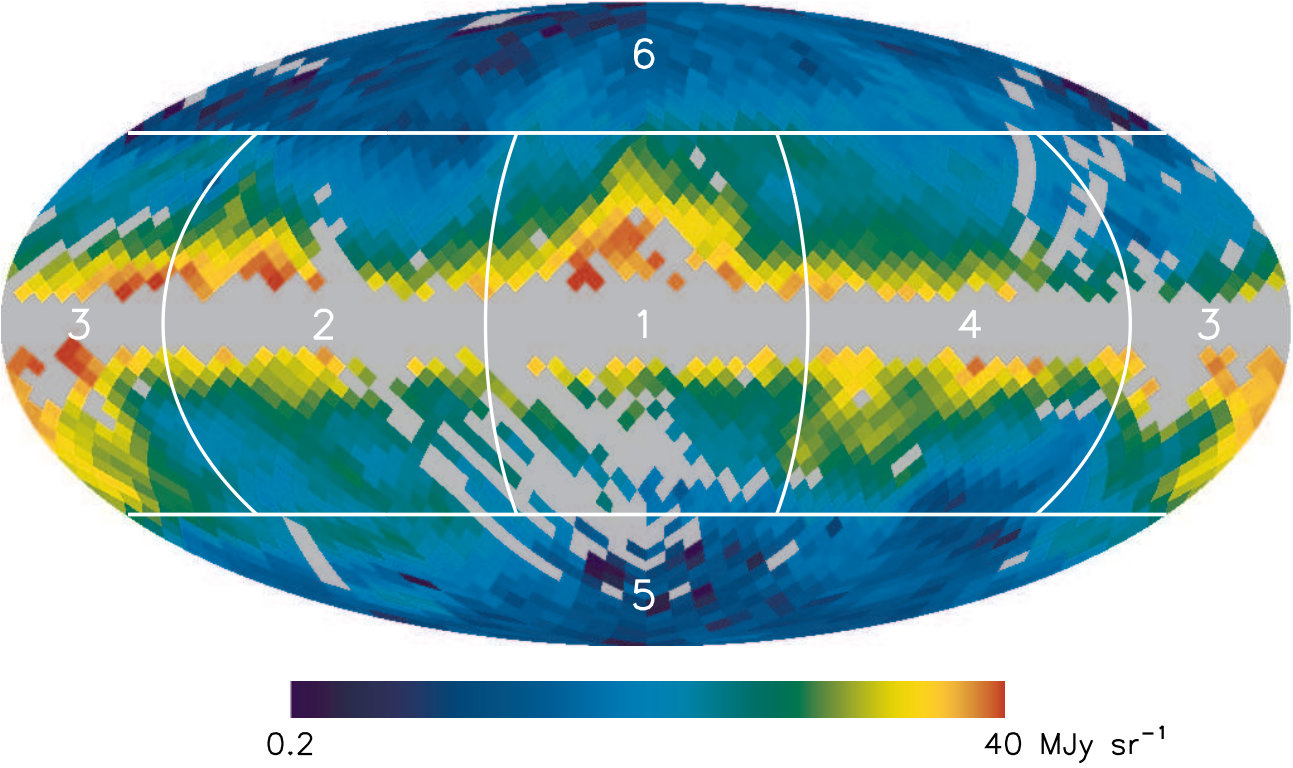}
\caption{A map of Galactic dust emission from FIRAS observations averaged over seven channels centered at 1200 GHz
(250 $\mu$m), on a logarithmic scale from 0.2 MJy sr$^{-1}$ to 40 MJy sr$^{-1}$.
Boundaries of the six large sky regions for which we form mean dust spectra are shown in white.
FIRAS pixels inside of the mask of FDS appear gray and are excluded from our analysis. The mask excludes
29\% of the sky.}
\end{figure}

\begin{figure}
\figurenum{3}
\epsscale{1.00}
\plotone{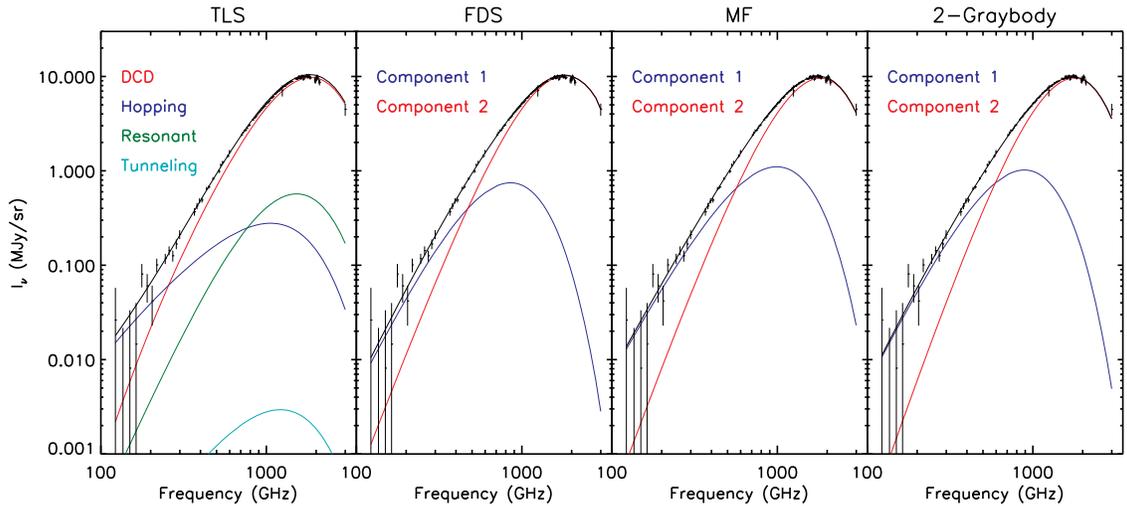}
\caption{The Galactic dust spectrum formed from FIRAS and DIRBE data for the sky region centered
at $l=90, b=0$, shown with results from fitting the two-level systems (TLS) model, FDS model,
MF model, and two-graybody model.  In each panel, the total model spectrum is shown as the black 
curve and individual model components are shown in color.}
\end{figure}

\begin{figure}
\figurenum{4}
\epsscale{1.00}
\plotone{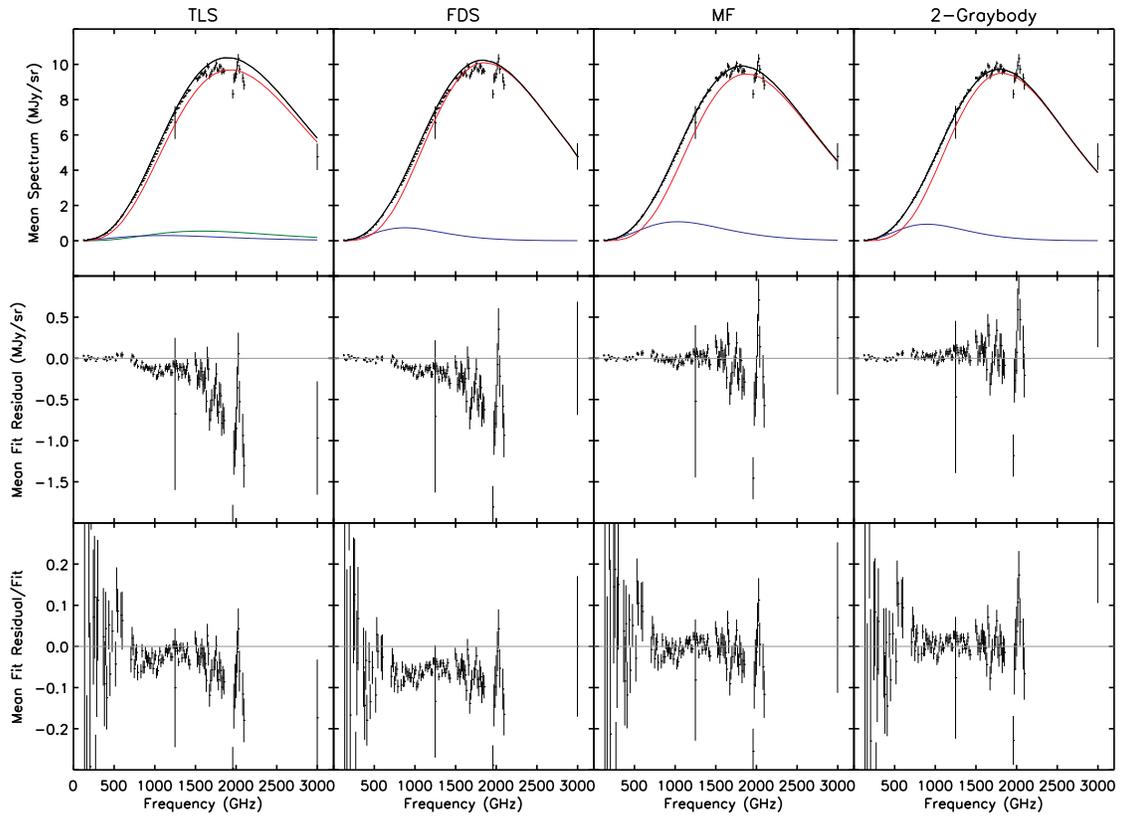}
\caption{Averages over the full sky outside masked regions of Galactic dust spectra and model fits (top row),
fit residuals (middle row), and fractional fit residuals (bottom row).
Results are shown for the TLS model fit, FDS model fit, MF model fit, and
two-graybody model fit.  The color coding for the model components
in the top row is the same as in Figure 3.}
\end{figure}

\clearpage

\begin{figure}
\figurenum{5}
\epsscale{1.00}
\plotone{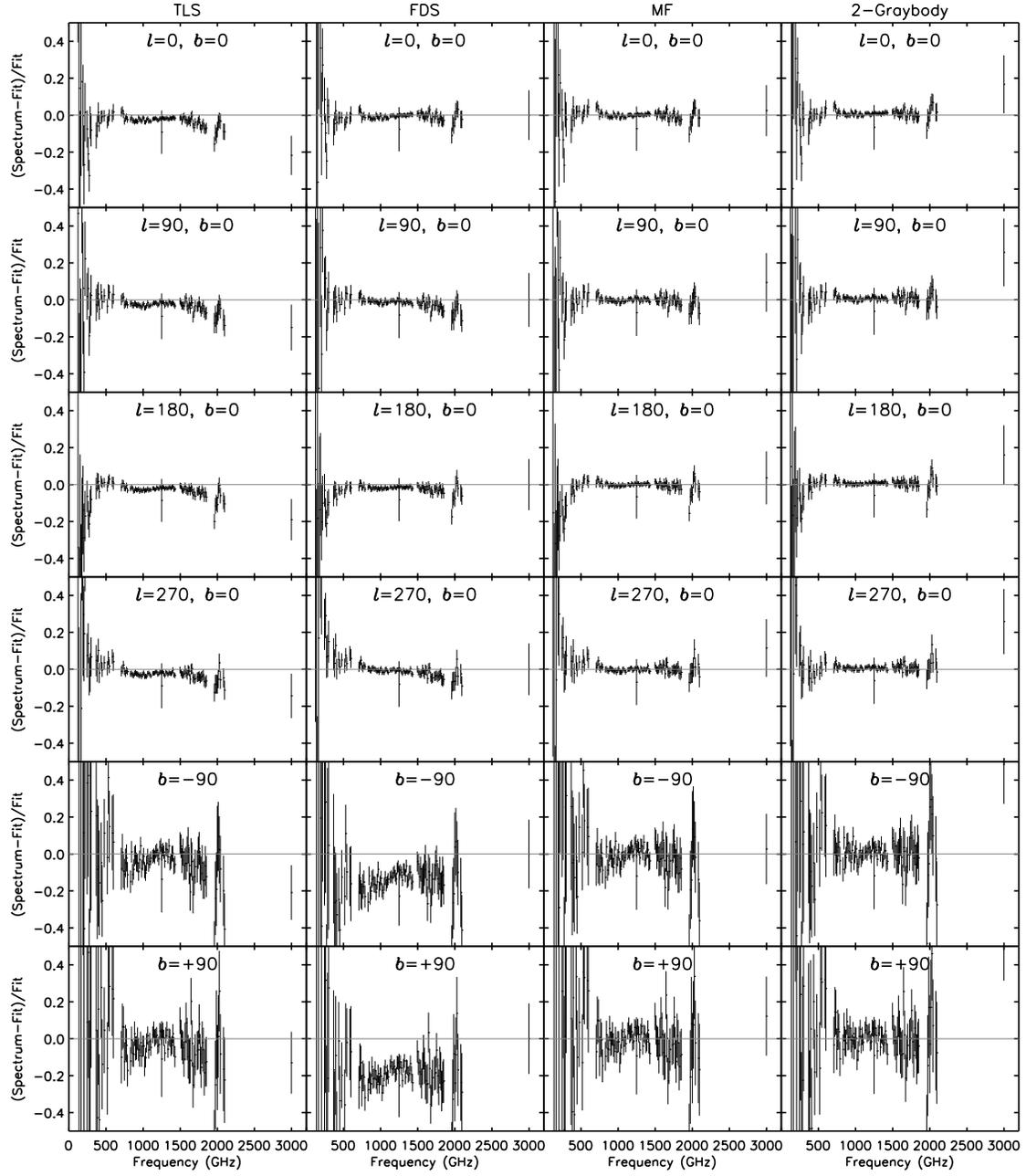}
\caption{Fractional residuals of the dust model fits for each sky region.
Results are shown for the TLS model fit, FDS model fit, MF model fit, and
two-graybody model fit.}
\end{figure}

\clearpage

\begin{figure}
\figurenum{6}
\epsscale{1.00}
\includegraphics[scale=0.75,angle=90]{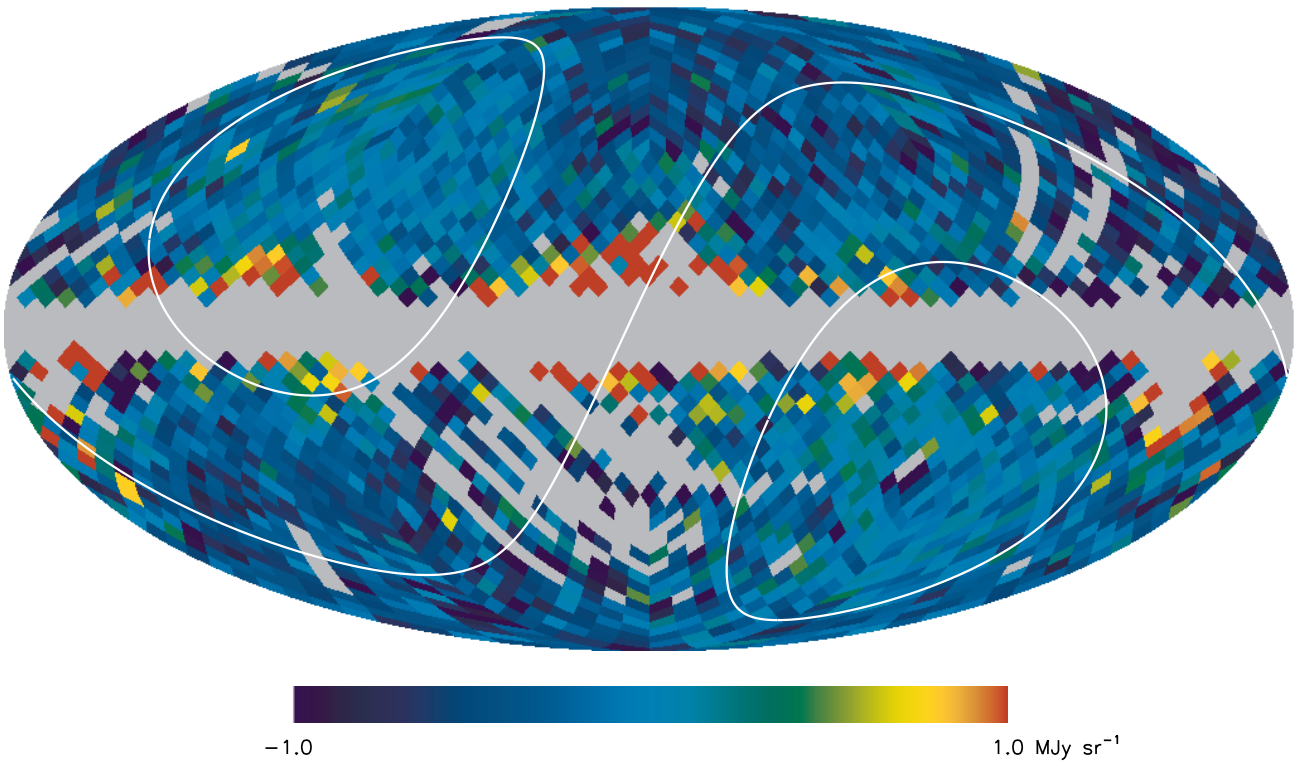}
\caption{Residual of the FDS model fit calculated at FIRAS resolution, averaged 
over seven channels centered at 1200 GHz (250 $\mu$m), on a linear scale from -1 to 1
MJy sr$^{-1}$.  Contours of ecliptic latitude equal to $-45\arcdeg, 0\arcdeg$, and 
$+45\arcdeg$ are shown in white. FIRAS pixels inside of the mask of FDS appear gray.}
\end{figure}

\begin{figure}
\figurenum{7}
\epsscale{1.00}
\plotone{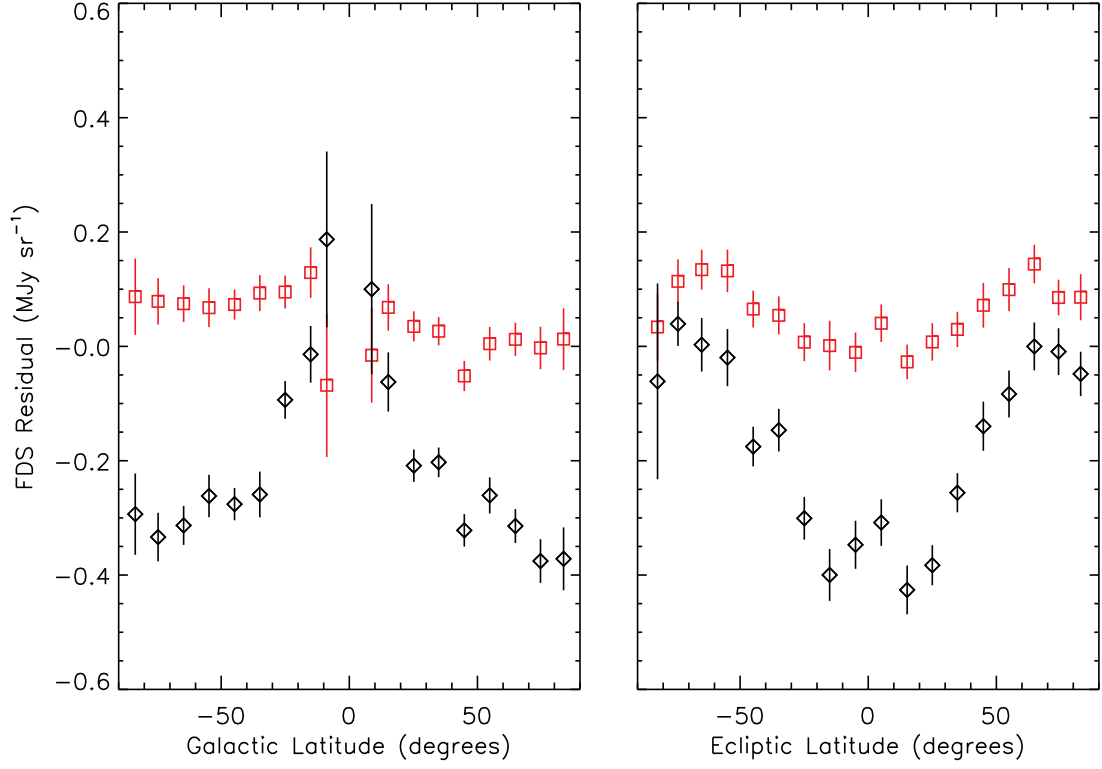}
\caption{Dependence of the FDS fit residual at 1200 GHz on Galactic latitude and 
ecliptic latitude.  The black diamonds show medians of the residual map shown in 
Figure 6 calculated over 10 degree bins in latitude. The red squares show results
from fitting a modified version of the FDS model that uses a rescaled 
SFD 100 $\mu$m map and a 100 $\mu$m/240 $\mu$m ratio map calculated from the 
{\it Planck} team thermal dust model.}
\end{figure}

\clearpage

\begin{deluxetable}{ccc}
\scriptsize
\tablewidth{0pt}
\tablecaption{Dust Model Fit Parameters}
\tablehead{
\colhead{Model} &
\colhead{Global Parameters} &
\colhead{Regional Parameters}
}
\startdata 
TLS Model  & Charge correlation length $l_c$         &  Dust temperature $T_d$ \\
           & Relative TLS amplitude $A$              &  Spectrum amplitude $a$ \\
           & Tunneling states parameter $c_{\Delta}$ &                         \\
           &                                         &                         \\
FDS Model  & Dust emissivity indices $\alpha_1, \alpha_2$ &      None          \\
           & Component 1 fractional power $f_1$           &                    \\
           & Cross-section parameter $q_1/q_2$            &                    \\ 
           &                                              &                    \\
MF Model   & Dust emissivity indices $\alpha_1, \alpha_2$ & Dust temperatures $T_2$ \\
           & Component 1 fractional power $f_1$           & Dust optical depths $\tau$ \\
           & Cross-section parameter $q_1/q_2$            &                    \\ 
           &                                              &                    \\
2-Graybody & Warm dust emissivity index $\alpha_2$        & Dust temperatures $T_1, T_2$ \\
  Model    &                                            & Dust optical depths $\tau_1, \tau_2$ \\
\enddata
\end{deluxetable}

\begin{deluxetable}{cccccc}
\scriptsize
\tablewidth{0pt}
\tablecaption{Dust Model Fits}
\tablehead{
\colhead{Model} &
\colhead{Global} &
\colhead{Fit $\chi^{2}$} &
\colhead{$\nu$} &
\colhead{P($\chi^{2} > $ Fit $\chi^{2}$)} &
\colhead{Fit $\chi^{2}_{\nu}$}\\
\colhead{ } &
\colhead{Parameters} &
\colhead{ } &
\colhead{ } &
\colhead{ } &
\colhead{ }
}
\startdata 
TLS Model  & $l_c = 15.6 \pm 7.8$ nm    &  690 & 663 & 0.231 & 1.04 \\
           & $A = 5.1 \pm 3.9$          &      &     &      &       \\
           & $c_{\Delta} = 1510 \pm 1310$ &    &     &      &       \\
           &                            &      &     &      &       \\
FDS Model  & $\alpha_1 = 1.69 \pm 0.10$ & 2049 & 662\tablenotemark{a} & $1.6 \times 10^{-141}$ & 3.10 \\
           & $\alpha_2 = 2.74 \pm 0.16$ &      &     &      &    \\
           & $f_1 = 0.037 \pm 0.014$    &      &     &      &    \\
           & $q_1/q_2 = 13.2 \pm 10.9$   &      &     &      &   \\ 
           &                            &      &     &      &    \\
MF Model  & $\alpha_1 = 1.33 \pm 0.12$ & 598 & 662 & 0.963 & 0.90 \\
           & $\alpha_2 = 2.91 \pm 0.45$ &      &     &      &    \\
           & $f_1 = 0.070 \pm 0.073$    &      &     &      &    \\
           & $q_1/q_2 = 11.0 \pm 13.7$   &      &     &      &    \\ 
           &                            &      &     &      &    \\
2-Graybody & $\alpha_1 = 1.7 $ (fixed) & 527 & 653 & 0.9998 & 0.81 \\
  Model     & $\alpha_2 = 3.20 $        &      &     &      &    \\
           &                            &      &     &      &   \\

\enddata
\tablenotetext{a}{The DIRBE 100 $\mu$m and 240 $\mu$m data points are not counted
in the number of degrees of freedom for the FDS model fit because the model uses 
DIRBE 100 $\mu$m and 240 $\mu$m maps in its calculation of the dust spectrum for
each region.}
\end{deluxetable}

\begin{deluxetable}{ccccccccc}
\scriptsize
\tablewidth{0pt}
\tablecaption{Dust Model Regional Parameters}
\tablehead{
\colhead{Model} &
\colhead{$\chi^{2}_{\nu}$} &
\colhead{Parameter} &
\colhead{Region 1} &
\colhead{Region 2} &
\colhead{Region 3} &
\colhead{Region 4} &
\colhead{Region 5} &
\colhead{Region 6}\\
\colhead{ } &
\colhead{ } &
\colhead{ } &
\colhead{(0, 0)\tablenotemark{a}} &
\colhead{(90, 0)} &
\colhead{(180, 0)} &
\colhead{(270, 0)} &
\colhead{(0, -90)} &
\colhead{(0, 90)}
}
\startdata 
TLS Model  & 1.04  & $T_{dust}$ (K)    & 19.5 & 18.0 &  17.9 & 18.6 & 19.5 & 19.8\\
           &      & $a$ ($10^{-8}$ cm) & 17  & 15   & 24    & 14 & 3.0  & 2.3 \\ 
           &      &               &      &      &       &      &      &     \\
FDS Model\tablenotemark{b} & 3.10 & $T_1$ (K) & 9.5  &  9.1 &  8.9  & 9.3  & 9.2  & 9.3 \\
           &      & $T_2$ (K)           & 16.2 & 15.7 & 15.3  & 15.9 & 15.8 & 16.0\\
           &      &                     &      &      &       &      &      &     \\
MF Model\tablenotemark{b} & 0.90  & $T_1$ (K) & 12.1 & 11.1 & 11.1 & 11.5 & 12.1 & 12.4 \\
           &      & $T_2$ (K)    & 15.8 & 14.9 & 14.8 & 15.3 & 15.9 & 16.1\\
           &      & $\tau/10^{-5}$ & 13.7 & 12.2 & 19.8 & 12.0 & 2.5 & 1.8 \\
           &      &                     &      &      &       &      &      &     \\

2-Gray Model & 0.81 & $T_1$ (K) & 9.6 & 9.2 & 9.5 & 9.0 & 6.5 & 6.7 \\
             &      & $T_2$ (K)         & 14.7 & 13.8 & 13.9 & 14.2 & 13.7 & 14.0 \\
             &      & $\tau_1/10^{-5}$  & 15.3 & 12.5 & 18.4 & 14.9 &  7.4 &  5.9 \\
             &      & $\tau_2/10^{-5}$  & 10.0 &  8.8 & 13.5 &  8.7 &  2.8 &  2.0 \\
           &      &                     &      &      &       &      &      &     \\
\enddata
\tablenotetext{a}{Galactic coordinates ($l,b$) of the region center.}
\tablenotetext{b}{The dust temperatures $T_1$ and $T_2$ for the FDS model and $T_1$ for the MF model are derived parameters. All
of the other parameters listed are fit parameters.}

\end{deluxetable}


\begin{references}

\reference{} Agladze, N. I., Sievers, A. J., Jones, S. A., et al. 1996, \apj, 462, 1026

\reference{} Anderson, P. W., Halperin, B. I., \& Varma, C. M., Philos. Mag., 25, 1

\reference{} Bennett, C.L., et al., 2013, ApJS., 208, 20B

\reference{} B{\"o}sch, M.A. 1978, Phys. Rev. Lett., 40, 879

\reference{} Boudet, N., Mutschke, H., Nayral, C., et al. 2005, \apj, 633, 272

\reference{} Bracco, A., Cooray, A., Veneziani, M., et al. 2011, MNRAS, 412, 1151

\reference{} Brodd, S., Fixsen, D. J., Jensen, K. A., Mather, J. C., \&
Shafer, R. A. 1997, {\it COBE} Far Infrared Absolute Spectrophotometer (FIRAS)
Explanatory Supplement, {\it COBE} Ref. Pub. No. 97-C (Greenbelt, MD: NASA/GSFC),
available in electronic form from http://lambda.gsfc.nasa.gov

\reference{} Compi{\`e}gne, M., Verstraete, L., Jones, A., et al., 2011, \aap, 525, A103

\reference{} Coupeaud, A., Demyk, K., Meny, C., et al. 2011, \aap, 535, A124

\reference{} D{\'e}sert, F.-X., Mac{\'i}as-P{\'e}rez, J. F., Mayet, F., et al. 2008, \aap, 481, 411

\reference{} Draine, B. T., and Hensley, B., 2013, \apj, 765, 159

\reference{} Draine, B. T., and Li, A., 2007, \apj, 657, 810

\reference{} Dupac, X., Boudet, N., Giard, M., et al. 2003, \aap, 404, L11

\reference{} Finkbeiner, D. P., Davis, M., \&  Schlegel, D. J., 1999, \apj, 524, 867 (FDS)

\reference{} Fixsen, D. J., Cheng, E. S., Gales, J. M., et al. 1996, \apj, 473, 576

\reference{} Fixsen, D. J., Weiland, J. L., Brodd, S., et al. 1997, \apj, 490, 482

\reference{Fix98} Fixsen, D. J., Dwek, E., Mather, J. C., Bennett, C. L.,
\& Shafer, R. A. 1998, \apj, 508, 123

\reference{Hau98} Hauser, M. G., et al. 1998, \apj, 508, 25

\reference{} Hinshaw, G., et al. 2009, \apjs, 180, 225

\reference{} Jones, A. P., Fanciullo, L., K{\"o}hler, M., et al., 2013, \aap, 558, A62

\reference{} Juvela, M., Montillaud, J., Ysard, N., \& Lunttila, T. 2013, \aap, 556, A63

\reference{} Kelly, B. C., Shetty, R., Stutz, A. M., et al. 2012, ApJ, 752, 55

\reference{} Kelsall, T., et al. 1998, \apj, 508, 44

\reference{} Liang, Z., Fixsen, D. J., \& Gold, B. 2012, MNRAS, submitted (arXiv:1201.0060)

\reference{} Meisner, A. M., and Finkbeiner, D. P. 2015, \apj, 798, 88

\reference{} Mennella, V., Brucato, J. R., Colangeli, L., et al. 1998, \apj, 496, 1058

\reference{} Meny, C., Gromov, V., Boudet, N., Bernard, J.-Ph., Paradis, D.,
\& Nayral, C. 2007, \aap, 468, 171

\reference{} Miville-Deschenes, M.-A. \&  Lagache, G. 2005, \apjs, 157, 302

\reference{} Nelder, J. A., \& Mead, R. 1965, Computer Journal, 7, 308

\reference{} Paradis, D., Bernard, J.-Ph., \& M{\'e}ny, C. 2009, \aap, 506, 745

\reference{} Paradis, D., Bernard, J.-Ph., M{\'e}ny, C., \& Gromov, V. 2011, \aap, 534, 118

\reference{} Paradis, D., Paladini, R., Noriega-Crespo, A., et al. 2012, \aap, 537, A113

\reference{} Paradis, D., Veneziani, M., Noriega-Crespo, A., et al. 2010, \aap, 520, L8

\reference{} Phillips, W. A. 1972, J. Low Temp. Phys., 7, 351

\reference{} Planck Collaboration Early XXIII, Ade, P. A. R., Aghanim, N., Arnaud, M., et al. 2011, \aap, 536, A23

\reference{} Planck Collaboration Early XXV, Abergel, A., Ade, P. A. R., Aghanim, N., et al. 2011, \aap, 536, A25

\reference{} Planck Collaboration Int. XVII, Abergel, A., Ade, P. A. R., Aghanim, N., Alves, M. I. R.,
et al., 2014, \aap, 566, 55

\reference{} Planck Collaboration Int. XXII, Ade, P. A. R., Alves, M. I. R., Aniano, G., Armitage-Caplan, C., et al.
2015, \aap, 576, A107

\reference{} Planck Collaboration 2013 Results XI, Abergel, A., Ade, P. A. R., Aghanim, N., et al. 2014, \aap, 571, A11

\reference{} Planck Collaboration 2015 Results VIII, Adam, R., Ade, P. A. R., Aghanim, N., et al. 2015, \aap,
submitted (arXiv:1502.01587)

\reference{} Planck Collaboration 2015 Results X, Adam, R., Ade, P. A. R., Aghanim, N., et al. 2015, \aap, 
submitted (arXiv:1502.01588)
 
\reference{} Reach, W. T., Dwek, E., Fixsen, D. J., et al., 1995, ApJ, 451, 188

\reference{} Schlegel, D. J., Finkbeiner, D. P., \& Davis, M. 1998, \apj, 500, 525 (SFD)

\reference{} Shetty, R. Kauffmann, J., Schnee, S., et al. 1996, \apj, 696, 2234

\reference{} Veneziani, M., Ade, P. A. R., Bock, J. J., et al. 2010, \apj, 713, 959

\reference{} Veneziani, M., Piacentini, F., Noriega-Crespo, A., et al. 2013, ApJ, 772, 56

\reference{} Wright, E. L., Mather, J. C., Bennett, C. L., et al., 1991, ApJ, 381, 200

\end{references}
\end{document}